\documentclass{elsart}
\usepackage{amssymb}
\usepackage[fleqn]{amsmath}
\usepackage{graphicx}
\usepackage{bigstrut}
\usepackage{graphics}
\usepackage{graphicx}
\usepackage{subfigure}
\usepackage{amssymb}
\usepackage{graphics}
\usepackage{graphicx}
\usepackage{color}

\newcommand{\der}[2]{\frac{\partial #1}{\partial #2}}

\def\p{\textbf{p}}
\def\q{\textbf{q}}

\newcommand{\me}[1]{\langle #1 \rangle}

\begin{document}
\begin{frontmatter}
\title{Two-fluid model of the Truncated Euler's Equations}
\author[ENS,GKmail]{Giorgio Krstulovic}
\author[ENS,MEBmail]{Marc-\'Etienne Brachet}

\address[ENS]{Laboratoire de Physique Statistique de l'Ecole Normale
Sup{\'e}rieure, \\
associ{\'e} au CNRS et aux Universit{\'e}s Paris VI et VII, 24 Rue
Lhomond, 75231 Paris, France}
\thanks[GKmail]{krstulov@lps.ens.fr}
\thanks[MEBmail]{brachet@lps.ens.fr}

\begin{abstract}
A phenomenological
two-fluid model of the (time-reversible)
spectrally-truncated $3D$ Euler equation
is proposed.
The thermalized small scales are first  shown to
be quasi-normal.
The effective viscosity and thermal diffusion are then determined,
using EDQNM closure and Monte-Carlo numerical computations.
Finally, the model is validated by comparing its dynamics with that of the original
truncated Euler equation.
\end{abstract}


\end{frontmatter}



\maketitle


\section{Introduction}\label{sec:Intro}
It is well-known that the (inviscid and conservative) truncated
Euler equation admits absolute equilibrium solutions with Gaussian
statistics, equipartition of kinetic energy among all Fourier modes
and thus an energy spectrum $E(k)\sim k^2$ \cite{Houches}. Recently,
Cichowlas et al.  \cite{CBDB-echel, TheseCichowlas} observed that
the Euler equation, with a very large (several hundreds) spectral
truncation wavenumber $k_{\rm max}$, has long-lasting transients
which behave just as those of high-Reynolds number viscous flow; in
particular they found an approximately $k^{-5/3}$ inertial range
followed by a dissipative range. How is such a behavior possible? It
was found that the highest-$k$ modes thermalize at first, displaying
a $k^2$ spectrum. Progressively the thermalized region extends to
lower and lower wavenumbers, eventually covering the whole range of
available modes. At intermediate times, when the thermalized regime
only extends over the highest wavenumbers, it acts as a thermostat
that pumps out the energy of larger-scale modes.
Note that similar $k^{-5/3}$/ $k^2$ spectra have already
been discussed in the wave turbulence literature ({\sl e.g.},\cite{Zakh})
and were more recently obtained, within a simple
differential closure, in connection with the Leith model of
hydrodynamic turbulence \cite{Colm}.

The purpose of the present work is to build a quantitative
two-fluid model for the relaxation of the $3 D$ Euler equation.
In section \ref{sec:Relaxation},
after a brief recall of basic definitions, the statistics of
the thermalized small scales are studied during relaxation.
They are shown to be quasi-normal.
Our new  two-fluid model, involving both an effective viscosity and a thermal diffusion,
is introduced in section \ref{sec:Two-fluid}. The effective diffusion laws are then determined,
using an EDQNM closure prediction and direct Monte-Carlo computations.
The model is then validated by comparing its predictions with the behavior of the original
truncated Euler equation.  Finally section \ref{sec:Conc} is our conclusion.

\section{Relaxation dynamics of truncated Euler
equations}\label{sec:Relaxation}
\subsection{Basic definitions}
The  truncated Euler equations (\ref{eq_discrt}) are classically
obtained \cite{Houches} by performing a Galerkin truncation ($\hat{v}(k)=0$
for  $sup_\alpha |k_\alpha | > k_{max}$) on the Fourier
transform  ${\bf v}({\bf x},t)=\sum {\bf \hat v}({\bf k},t) e^{i
{\bf k}\cdot {\bf x}}$ of a spatially periodic velocity field
obeying the (unit density) three-dimensional incompressible Euler
equations,
${\partial_t {\bf v}}  + ({\bf v} \cdot \nabla) {\bf v} =- \nabla p   ~,  $
 $\nabla  \cdot {\bf v} =0 $.
This procedure yields the following finite
system of ordinary differentials equations for the complex variables
${\bf \hat v}({\bf k})$ (${\bf k}$ is a 3 D vector of relative
integers $(k_1,k_2,k_3)$ satisfying $\sup_\alpha |k_\alpha | \leq
k_{\rm max}$)
\begin{equation}
{\partial_t { \hat v}_\alpha({\bf k},t)}  =  -\frac{i} {2} {\mathcal
P}_{\alpha \beta \gamma}({\bf k}) \sum_{\bf p} {\hat v}_\beta({\bf
p},t) {\hat v}_\gamma({\bf k-p},t) \label{eq_discrt}
\end{equation}
where ${\mathcal P}_{\alpha \beta \gamma}=k_\beta P_{\alpha
\gamma}+k_\gamma P_{\alpha \beta}$ with $P_{\alpha
\beta}=\delta_{\alpha \beta}-k_\alpha k_\beta/k^2$ and the
convolution in (\ref{eq_discrt}) is truncated to $\sup_\alpha
|k_\alpha | \leq k_{\rm max}$, $\sup_\alpha |p_\alpha | \leq k_{\rm
max}$ and $\sup_\alpha |k_\alpha-p_\alpha | \leq k_{\rm max}$.

This time-reversible system exactly conserves the kinetic
energy $E=\sum_{k}E(k,t)$, where the energy spectrum $E(k,t)$ is
defined by averaging ${\bf \hat v}({\bf k'},t)$ on spherical shells
of width $\Delta k = 1$,
\begin{equation}
E(k,t) = {\frac1 2} \sum_{k-\Delta k/2< |{\bf k'}| <  k + \Delta
k/2} |{\bf \hat v}({\bf k'},t)|^2 \, . \label{eq_energy}
\end{equation}
\subsection{Small Scales Statistics}
Perhaps the most striking result of Cichowlas
et al.  \cite{CBDB-echel} was the spontaneous generation
of a (time dependent)
minimum of the spectrum  $E(k,t)$ at wavenumber $k_{\rm th}(t)$
where the scaling law $E(k,t)=c(t) k^2$ starts.
Thus, the energy dissipated from large scales into the time dependent statistical equilibrium is given by
\begin{equation}
{E}_{\rm  th}(t) =  \sum_{k_{\rm th}(t) <  k } E(k,t) ~.
\label{Th_energy}
\end{equation}
In this section we use
the so-called Taylor-Green \cite{TG1937} initial condition to (\ref{eq_discrt}):
the single--mode Fourier transform of  ${u}^{\rm
TG}=\sin{x}\cos{y}\cos{z}$, ${v}^{\rm TG}=-{u}^{\rm TG}(y,-x,z)$,
${w}^{\rm TG}=0$.

In order to separate the dynamics of large-scale ($k<k_{\rm th}$)
and the statistics of small-scales ($k>k_{\rm th}$)
we define the low and high-pass filtered fields
\begin{eqnarray}
  f^<(\textbf{r}) &=& \sum\limits_k F(\textbf{k})\hat{f}_\textbf{k} e^{i\textbf{k}\cdot\textbf{r}} \label{eq:Filt1}\\
  f^>(\textbf{k}) &=& 1-f^<(\textbf{r}) \label{eq:Filt2}
\end{eqnarray}
where  $f(\textbf{r})$ is an arbitrary field  and $\hat{f}_k$  its
Fourier transform; we have chosen
$F(\textbf{k}=\frac{1}{2}\left(1+\tanh{\left[\frac{|k|-k_{\rm
th}}{\Delta k}\right]}\right)$, with ${\Delta k}=1/2$.

\begin{figure}[h!]
\begin{center}
  \includegraphics[width=0.4\textwidth]{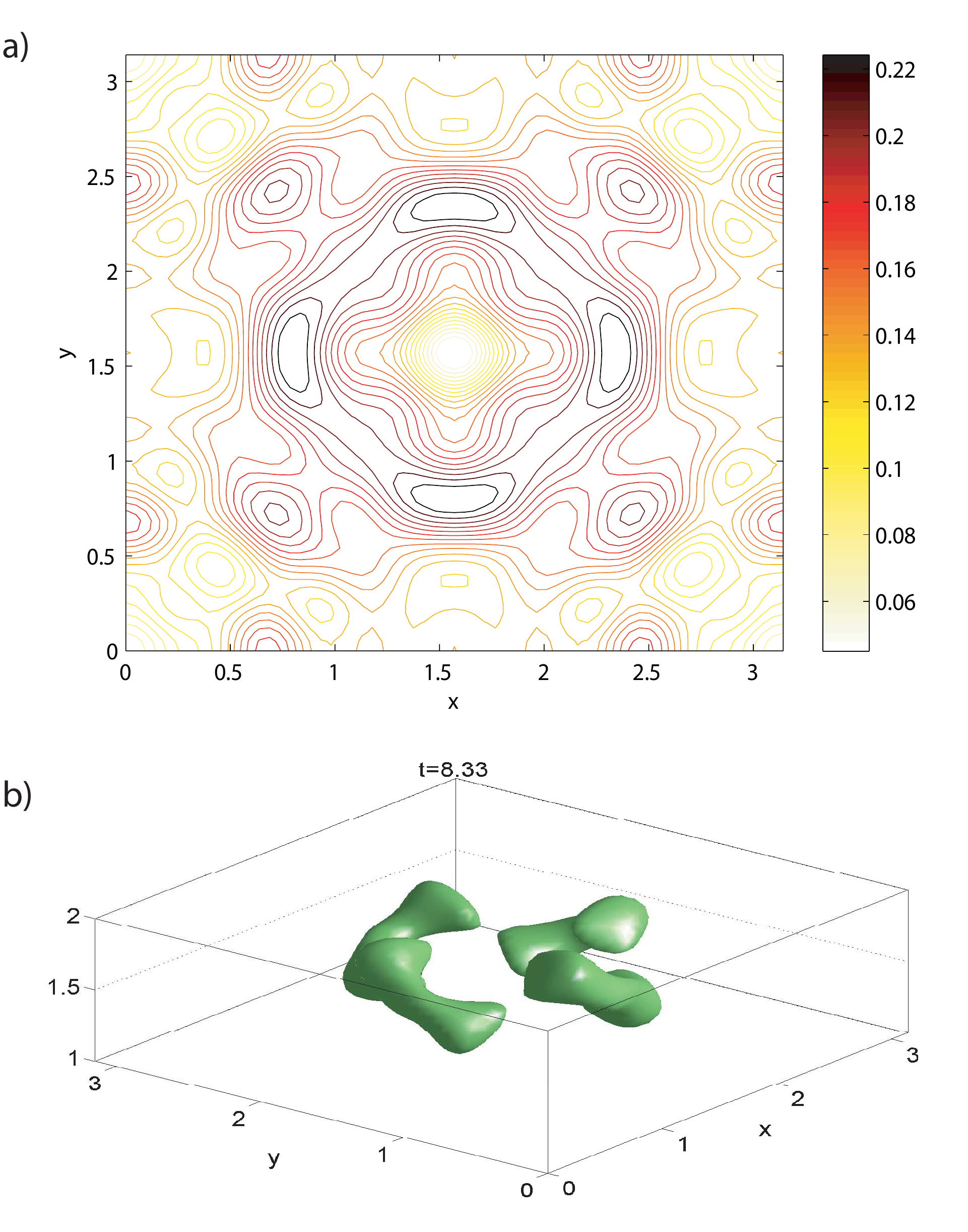}
  \caption{Cut at $z=\frac{\pi}{2}$ of $Q$ (a) and the isosurface
 $Q(r)=0.8Q_{max}=0.42$ (b). }\label{FigTemp}
\end{center}
\end{figure}

This filter allows us to define the large-scale velocity field ${\bf
v}^<$ and the spatially dependent thermalized energy (or heat)
associated to quasi-equilibrium. Using the trace of the Reynold's
tensor \cite{K-eps} , $R_{ij}=\frac{1}{2}(v_i^>v_j^>)^<$, we  define
the local heat as
\begin{equation}
Q(\textbf{r})=\frac{1}{2}\left[({\bf v}^>)^2\right]^<(\textbf{r}).
\label{eq:Q}
\end{equation}
By construction of the filters, (\ref{eq:Filt1}-\ref{eq:Filt2}) the
heat spatial average is equal to the dissipated energy
(\ref{Th_energy}) $<Q(r)>={E}_{\rm  th}$. Fig. \ref{FigTemp}a shows
a $2D$ cut of the heat $Q$ on the surface $z=\frac{\pi}{2}$, where a
cold zone is seen to be present at the center of the impermeable box
($x=[0,\pi]$, $y=[0,\pi]$, $z=[0,\pi]$). An isosurface of the
hottest zones is displayed on Fig. \ref{FigTemp}b. Is is apparent on
both figures that $Q{\bf (r)}$ is not spatially homogeneous.
\subsection{Heat diffusion}
The simplest quantities to study in order to quantify the evolution
of $Q$, are the spatial average $Q(t)=\langle
Q(\textbf{r},t)\rangle$ and the root mean square variation $\Delta
Q=\sqrt{\langle (Q^2-\langle Q\rangle^2)\rangle}$.
\begin{figure}[h!]
\begin{center}
  \includegraphics[width=0.45\textwidth,height=5cm]{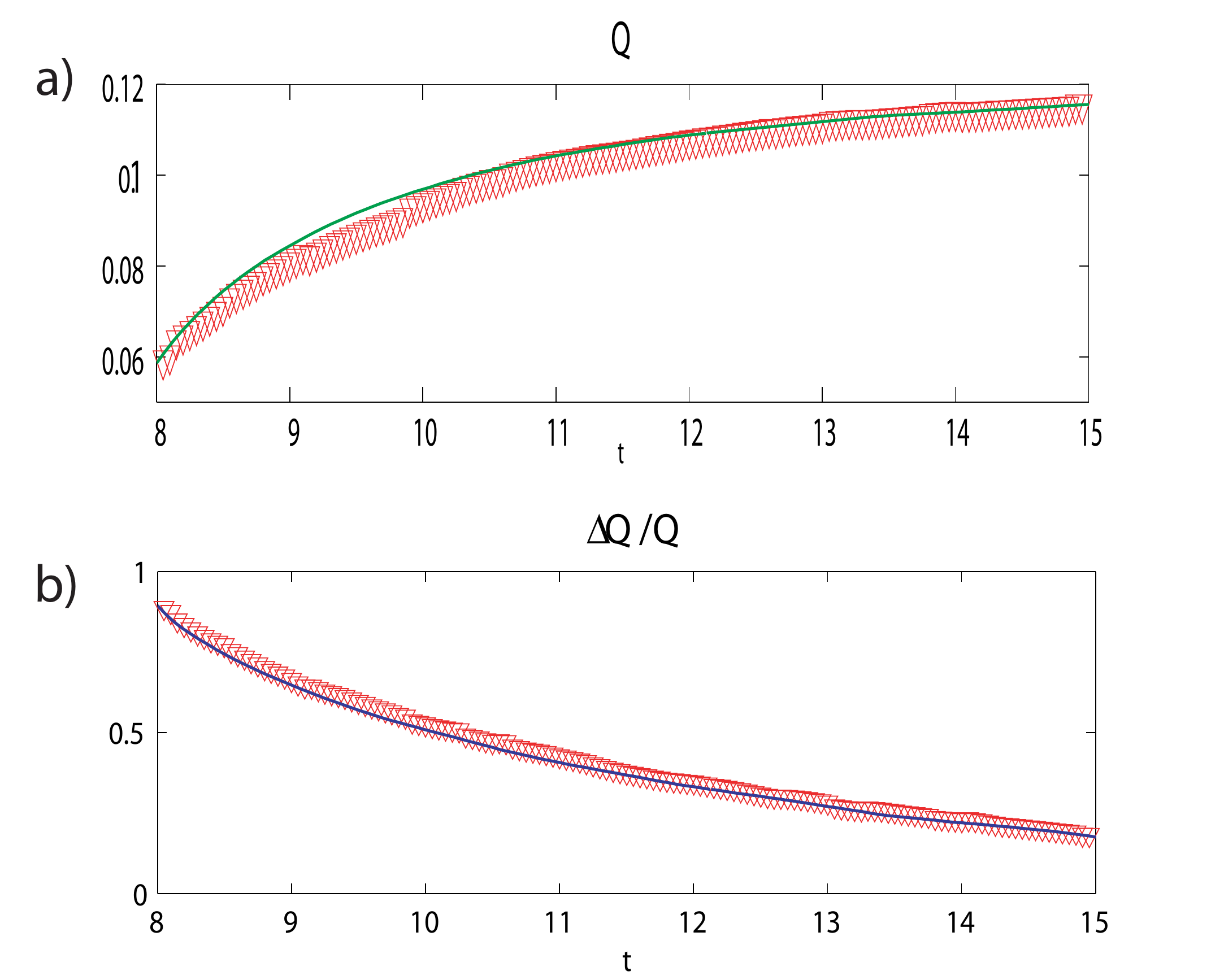}
  \caption{Plots of $Q(t)$ (a) and $\Delta Q(t)/ Q(t)$ (b); solid lines are the results of the two-fluid model (see section \ref{sec:Two-fluid}).}\label{FigTempVst}
\end{center}
\end{figure}
These quantities are shown in figure \ref{FigTempVst}, where that
the mean heat is seen to increases in time, due to the energy coming
from the large eddies, as was shown precedently in
\cite{CBDB-echel}. The relative fluctuation $\Delta Q/Q$ is seen to
decrease from $0.9$ to $0.2$.

\begin{figure}[h!]
\begin{center}
  \includegraphics[width=0.71\textwidth]{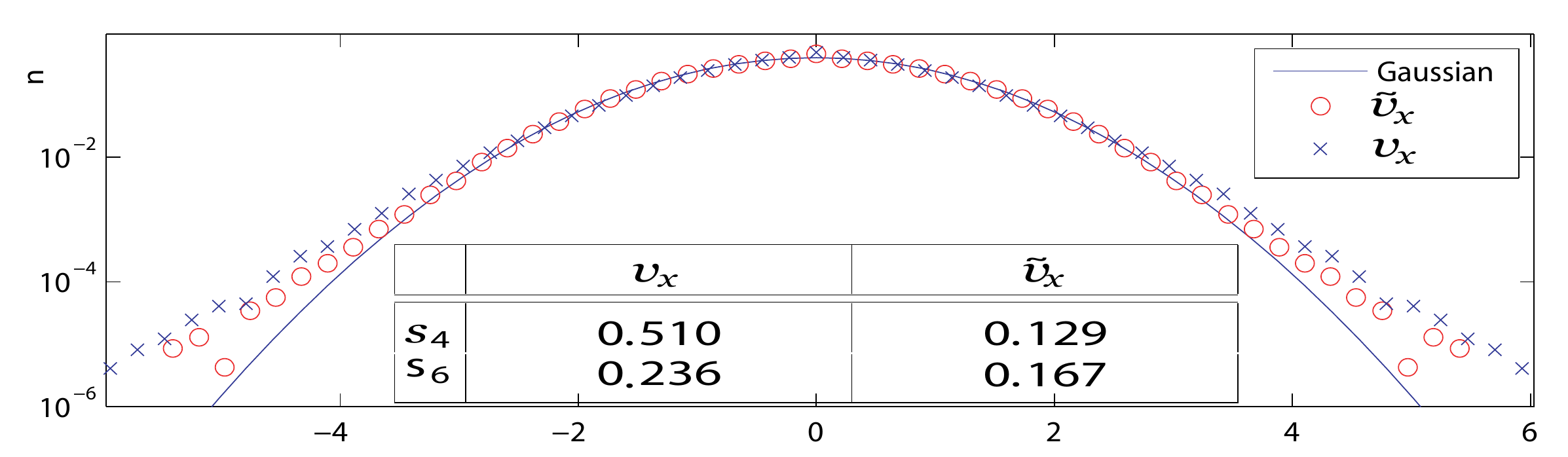}
  \caption{Histogram of $v^>_x$ and $\tilde {v}^>_x$ and normalized cumulant $s_4$ and  $s_6$ (odd cumulants vanish because of symmetries).}\label{FigHistoV}
\end{center}
\end{figure}

The next natural question is related to the statistical distribution
of the small eddies $v^>$: are they approximately Gaussian, like an
absolute equilibrium? A histogram of $v^>_x$ is shown in figure
\ref{FigHistoV}
.
As the heat is not homogeneous, we also computed the histogram of
the normalized field $\tilde {v}^>_x=v^>_x/\sqrt{Q}$ which seems to
better obey Gaussian statistics as can be seen on figure
\ref{FigHistoV} and comparing the firsts normalized cumulant
$s_n=\frac{c_n}{\sqrt{c_2^n}}$ ($c_n$ is the cumulant of order $n$)
in the table.

\section{Two-fluid Model}\label{sec:Two-fluid}
We now introduce our phenomenological two-fluid model of
the truncated Euler equation. One of the fluids describes the
large scale velocity field and the other represents the thermalized
high-wavenumber modes described by a temperature field $T=Q/c$
($c$ is the specific heat, explicitly given by $c=8k_{\rm{max}}^3$).
This model is somewhat analogous to Landau's standard
two-fluid model of liquid Helium at finite temperature $T$ where there
is a natural cut-off wavenumber for thermal excitations:
the classical-quantum crossover wavenumber $k_{\rm max}$
given by $\hbar  k_{\rm max} c_S = k_{B}T$ ($c_S$ is the sound velocity
and $k_{B}$ Boltzmann's constant). In Landau's model $k_{\rm max}$
is temperature dependent and the specific heat $c$ is proportional to $T^3$.
In constrast, $k_{\rm{max}}$ and the specific heat are constant in our model that reads:
\begin{eqnarray}
  \partial_t v_i^<+v_j^< \partial_j v_i^< &=& -\partial_i \tilde{p} +\partial_j\sigma'_{ij}\label{Eqdeuxfluid1}\\
  \partial_i v_i^< &=& 0\label{Eqdeuxfluid2}\\
   \partial_t T+v_j^< \partial_j T&=& \mathcal{D}T+\frac{1}{2c}\left(\partial_j v_i^<+\partial_i
   v_j^<\right)\sigma'_{ij}\label{Eqdeuxfluid3}
\end{eqnarray}
where
\begin{eqnarray}
\sigma'_{ij}&=&\mathcal{F}^{-1}[\nu_{\rm{eff}}({
k})(i k_i\hat{v}_j^<+ i k_j\hat{v}_i^< )] \label{Eqdeuxfluid4}\\
\mathcal{D}T&=&\mathcal{F}^{-1}[- k^2
D_{\rm{eff}}(k)\mathcal{F}[T]]\label{Eqdeuxfluid5}
  \end{eqnarray}
and $\mathcal{F}[\cdot]$ denotes the Fourier transform.
$\sigma'_{ij}$ is a generalized form of the standard viscous strain
tensor \cite{LandauFluides} . The precise form of the anomalous
diffusion terms $\nu_{\rm{eff}}$ and $D_{\rm{eff}}$ will be
determined below, in sections \ref{subsec:EDQNM} and
\ref{subsec:NumDiff}.

The advection terms in equation (\ref{Eqdeuxfluid1}) are readily
obtained from the Reynolds equations for the filtered velocity  by
remarking that the diagonal part of the Reynolds stress can, because
of incompressibility, be absorbed in the pressure. Equation
(\ref{Eqdeuxfluid4}) represents a simple model of the traceless part
of the Reynolds tensor \cite{K-eps}. In the same vein, the advection
terms in equation  (\ref{Eqdeuxfluid3}) are readily obtained
together with higher-order moments (see equation (1) of reference
\cite{Launder}). The dissipation and source terms in
(\ref{Eqdeuxfluid3}) are thus simple models of the higher-order
moments. It is easy to show that in the present model $\langle
\frac{1}{2}{\bf v}^2+ c T\rangle$ is conserved,  corresponding to
the energy conservation in the truncated Euler equation.

As the fluctuations $\Delta Q/Q$ are small (see above) we will
furthermore assume that $\nu_{\rm{eff}}$ and $D_{\rm{eff}}$ only
depend on $<Q>=E_{\rm th}$. Thus the evolution of the filtered
velocity ${\bf v^<}$ is independent of the fluctuations $\Delta Q$.
As $[E_{\rm th}]=L^2 T^{-2}$, simple dimensional analysis yields the
following form for the function $\nu_{\rm{eff}}$ and $D_{\rm{eff}}$:
\small
\begin{equation}
\nu_{\rm{eff}}=\frac{\sqrt{E_{\rm{th}}}}{k_{\rm{max}}}f\left(\frac{k}{k_{\rm{max}}},\frac{k_0}{k_{\rm{max}}}\right)\,;
   D_{\rm{eff}}=\frac{\sqrt{E_{\rm{th}}}}{k_{\rm{max}}}\Psi\left(\frac{k}{k_{\rm{max}}},\frac{k_0}{k_{\rm{max}}}\right)\label{diffdimana}
\end{equation}\normalsize
where $k_0=2 \pi/L_{\rm p}$  the smallest nonzero wavenumber
($L_{\rm p}$ is the periodicity length, $2 \pi$ in the present
simulations).
\subsection{EDQNM Determination of viscosity\label{subsec:EDQNM}}
An analytical determination of function $\nu_{\rm{eff}}$ is possible
using the eddy-damped quasi-Markovian theory
(EDQNM)\cite{OrzagAnalytTheo}.
It is known that this model
well reproduces the dynamics of truncated Euler Equation,
including the $k^{-5/3}$ and $k^2$ scalings
and the relaxation to equilibrium \cite{BosBertoglioEDQNM}.

The EDQNM closure furnishes an integro-differential equation for
the spectrum $E(k,t)$:
\begin{equation}\label{eq_EDQNM}
    \der{E(k,t)}{t}=T_{NL}(k,t)
\end{equation}
where the nonlinear transfer $T_{NL}$ is modeled as
\begin{eqnarray}
\nonumber T_{NL}(k,t)&=&\int\int\limits_\triangle\Theta_{kpq}(xy+z^3)[k^2pE(p,t)E(q,t)\\
&-&p^3E(q,t)E(k,t)]\frac{dp\,dq}{pq} \label{eq_Tnl}.
\end{eqnarray}
In (\ref{eq_Tnl}) $\triangle$ is a strip in $p,q$ space such that
the three wavevectors $\textbf{k}$, $\p$, $\q$ form a triangle. $x$,
$y$, $z$, are the cosine of the angles opposite to $\textbf{k}$,
$\textbf{p}$, $\textbf{q}$. $\Theta_{kpq}$ is a characteristic time
defined as
\begin{equation}\label{eq_Theta}
    \Theta_{kpq}=\frac{1-\exp{(-(\eta_k+\eta_p+\eta_q)t)}}{\eta_k+\eta_p+\eta_q}
\end{equation}
and the eddy damped $\eta$ is defined as
\begin{equation}\label{eq_eta}
    \eta_k=\lambda\sqrt{\int\limits_0^k s^2E(s,t)\,ds}.
\end{equation}
Classically $\lambda=0.36$ and the truncation is imposed omitting
all interactions involving waves numbers larger than $k_{\rm{max}}$
in (\ref{eq_Tnl}).

A simple and important stationary solution of (\ref{eq_EDQNM}) is the
absolute equilibrium with equipartition of the kinetic
energy and corresponding spectrum $E(k)\sim k^2$.

To compute the EDQNM effective viscosity
$\nu_{\rm{eff}}$ we consider an absolute equilibrium with a small
perturbation added in the mode of wavenumber $k_{pert}$ and study
the relaxation to equilibrium.
The corresponding ansatz is
$E(p,t)=\frac{3
E_{\rm{th}}}{k_{\rm{max}}^3}p^2+\gamma(t)\delta(p-k_{\rm{pert}})$
and we suppose $E_{\rm{th}}\gg\gamma$,
so that the total energy is almost constant and equal to $E_{\rm{th}}$.

Using the long time limit of (\ref{eq_Theta}) and expanding the
EDQNM transfer (\ref{eq_Tnl}) to first order in $\gamma$ yields for
the delta containing part, after a lengthy but straightforward
computation:
\begin{equation}\label{eq_Tnlcal}
T_{NL}(k,t)=-\gamma(t)\delta(k-k_{\rm{pert}})k^2\frac{\sqrt{E_{\rm{th}}}}{k_{\rm{max}}}\frac{\sqrt{30}}{\lambda}I\left(\frac{k}{k_{\rm{max}}}\right)
\end{equation} where $I$ is  given by the explicit integral
\begin{equation*}
I(x)=\sqrt{x}\int\limits_1^{\frac{2-x}{x}}\int\limits_{-1}^1\frac{(p^2-1)(1-q^2)(q^2+p^2(1+2q^2)))}{(p^2-q^2)(2^\frac{5}{2}+((p-q)^\frac{5}{2}+(p+q)^\frac{5}{2}))}\,dq\,dp
\end{equation*}
Using (\ref{eq_EDQNM}), (\ref{eq_Tnlcal}) and the basic definition of the two-fluid model
(\ref{Eqdeuxfluid1}-\ref{Eqdeuxfluid5}), we obtain
\begin{equation}\label{eq_nueff}
    \nu_{\rm{eff}}(k)=\frac{\sqrt{E_{\rm{th}}}}{k_{\rm{max}}}\frac{\sqrt{30}}{2 \lambda}I\left(\frac{k}{k_{\rm{max}}}\right).
\end{equation}
The function $f(x=\frac{k}{k_{\rm{max}}},0)$  in (\ref{diffdimana})
is thus given by
\begin{equation}\label{eq_fnueff}
f(x,0)=\frac{\sqrt{30}}{2\lambda}I(x).
\end{equation}
In the limit $x \to 0$, it is simple to show that $f$ has a finite value
$f(0,0)=\frac{7}{\sqrt{15}\lambda}$.
Thus the EDQNM prediction in the small $k/k_{\rm{max}}$ limit is
\begin{equation}\label{eq_nuEDQNM}
\nu_{\rm{eff}}=\frac{\sqrt{E_{\rm{th}}}}{k_{\rm{max}}}\frac{7}{\sqrt{15}\lambda},
\end{equation}
with $\frac{7}{\sqrt{15}\lambda}=5.021$ for the classic value of
$\lambda=0.36$. This asymptotic value can also be obtained from the
EDQNM eddy viscosity expression calculated by Lesieur and Schertzer
\cite{LesieruSchertezerEDQNMExpa} using an energy spectrum $E(k)\sim
k^2$.
\subsection{Monte-Carlo determination of viscosity and thermal diffusion \label{subsec:NumDiff}}
In order to numerically determine the effective viscosity
$\nu_{\rm{eff}}(k)$ of the two-fluid model, we
use a general-periodic code to study the relaxation
of an absolute equilibrium perturbed by adding a stationary solution
of the Euler equation.
We thus consider the initial condition
\begin{eqnarray}
  u &=& \cos{kx}\sin{ky}+u_{\rm{eq}}\label{rot+eq1} \\
  v &=& -\sin{kx}\cos{ky}+v_{\rm{eq}}\label{rot+eq2} \\
  w &=& w_{\rm{eq}}\label{rot+eq3}
\end{eqnarray}
where the (solenoidal and Gaussian) absolute equilibrium velocity
field satisfies
$\me{u_{\rm{eq}}^2+v_{\rm{eq}}^2+w_{\rm{eq}}^2}=2E_{\rm{th}}$.

The resulting amplitude of
the rotation in (\ref{rot+eq1}-\ref{rot+eq3}) is found, after a
short transient, to decay exponentially in time.
\begin{figure}[h!]
\begin{center}
  \includegraphics[width=0.8\textwidth]{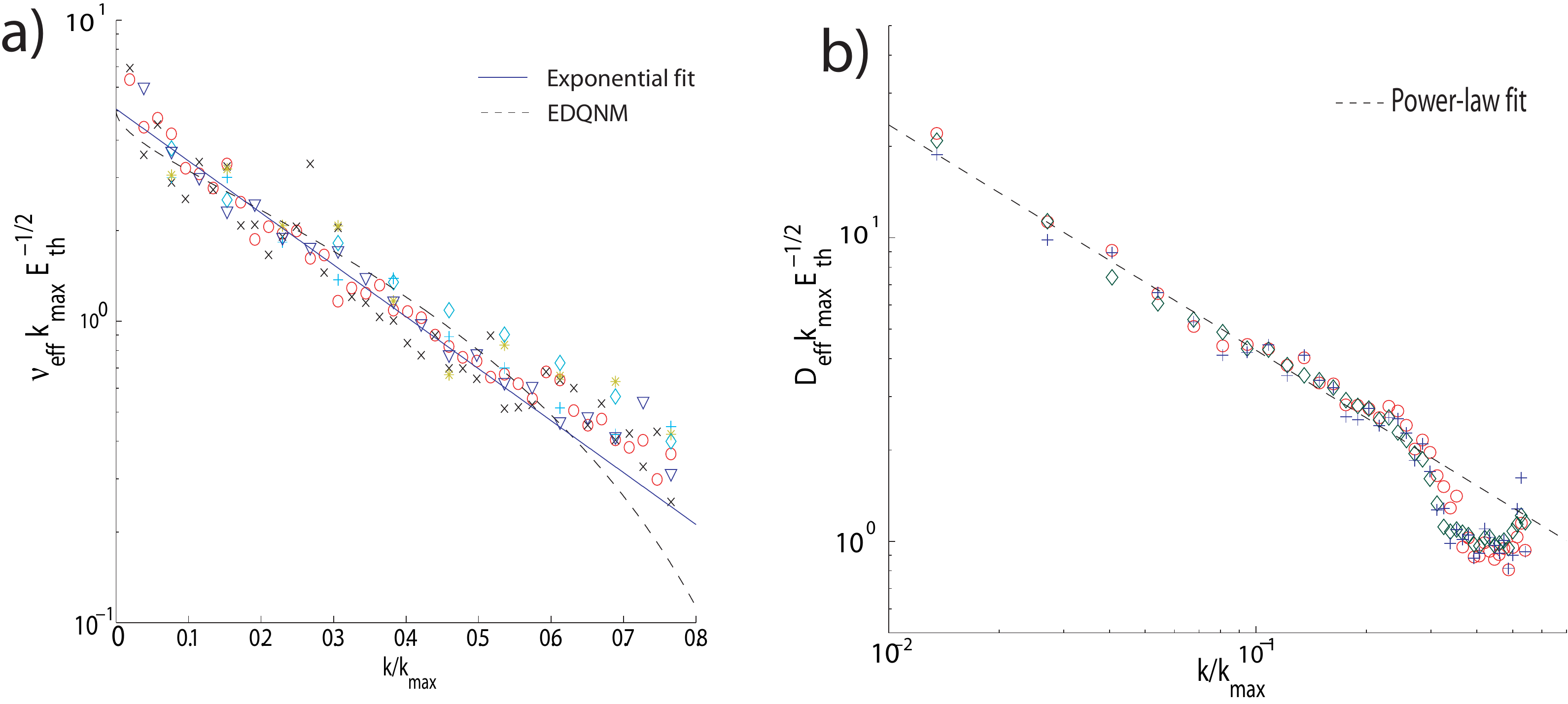}
  \caption{Effective viscosity $\nu_{\rm{eff}}$ (a) and thermal
diffusivity $D_{\rm{eff}}$ (b) determined by Monte Carlo computations performed at different values of $E_{\rm th}$ and $k_{\rm{max}}$ (see text).}\label{Figdiff}
\end{center}
\end{figure}
The function $\nu_{\rm{eff}}(k)$ is then obtained by finding the halving  time
$\tau_k$, for which $\hat{v}_\alpha({\bf
k},t_0+\tau_k)=\hat{v}_\alpha({\bf k},t_0)/2$,
with $t_0$ chosen larger than
the short transient time.
The effective dissipation thus reads
\begin{equation}
\nu_{\rm{eff}}(k)=\log{2}/(k^2\tau_k).\label{eq:nueffNum}
\end{equation}

The values of  $\nu_{\rm{eff}}(k)k_{\rm{max}}/\sqrt{E_{\rm{th}}}$
are shown in figure (\ref{Figdiff}a) for different values of
$E_{\rm{th}},k,k_{\rm{max}}$. A very good agreement with the EDQNM
prediction is observed. Note that there is not dependence in the
dimensionless parameter $k_0/k_{\rm{max}}$ (see eq.
(\ref{diffdimana})).

An exponential fit of all data in figure \ref{Figdiff}a gives
\begin{equation}
\nu_{\rm{eff}}=5.0723\frac{\sqrt{E_{\rm
th}}}{k_{\rm{max}}}e^{-3.97k/k_{\rm{max}}}.\label{eq:FitNueff}
\end{equation}
Note that the limit $k/k_{\rm{max}}\to 0$ is consistent with the
EDQNM prediction (\ref{eq_nuEDQNM}).

Another simple numerical experiment can be used to
characterize the thermal diffusion:  the relaxation of a
spatially-modulated
\textit{pseudo}-equilibrium  defined by
\begin{equation}
\left\langle u^2+v^2+w^2\right\rangle=2E_{\rm th}+2\epsilon
\cos{(kx)}\label{tempmod}
\end{equation}
with $\epsilon< E_{\rm th}$.

An $x$-dependent temperature can be recovered by averaging
$u^2+v^2+w^2$ over $y$ and $z$. Numerical integration of the
truncated Euler equation with the initial condition (\ref{tempmod})
produces an amplitude $\epsilon$ that decays exponentially, as in
the case studied for the determination of effective viscosity. The
thermal diffusivity $D_{\rm{eff}}$ is determined in the same way as
in eq. (\ref{eq:nueffNum}) and the corresponding data is shown in
figure \ref{Figdiff}b. A power-law fit gives
\begin{equation}
D_{\rm{eff}}=0.7723\frac{\sqrt{E_{\rm
th}}}{k_{\rm{max}}}(k/k_{\rm{max}})^{-0.74}\label{eq:FitDeff}.
\end{equation}
The negative exponent in (\ref{eq:FitDeff}) is characteristic of
hypodiffusive processes.
\begin{figure}[h!]
\begin{center}
  \includegraphics[width=0.6\textwidth,height=4.5cm]{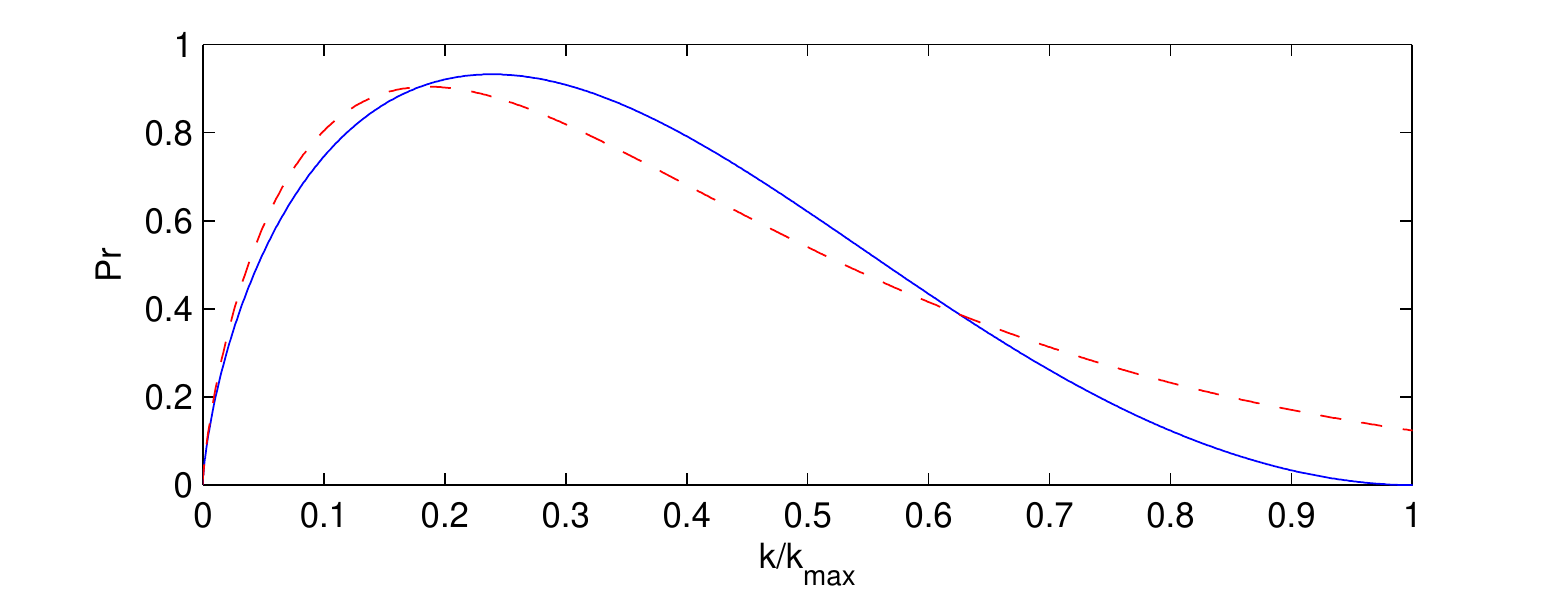}
  \caption{Effective Prandtl number $P_{\rm{eff}}=\nu_{\rm{eff}}/D_{\rm{eff}}$.}\label{FigPreff}
\end{center}
\end{figure}

We can define an effective Prandtl number as the ratio
$P_{\rm{eff}}(k)=\nu_{\rm{eff}}(k)/D_{\rm{eff}}(k)$. The Prandtl
number is plotted in figure \ref{FigPreff}, where the solid blue
line is obtained using the EDQNM prediction (\ref{eq_nuEDQNM}) and
the fit (\ref{eq:FitDeff}) and the dashed red line is obtained using
the fits (\ref{eq:FitNueff}) and (\ref{eq:FitDeff}). Note that the
Prandtl vanishes in the the small $k/k_{\rm{max}}$ limit and
verifies $P_{\rm{eff}}<1$ for all wavenumbers.
\subsection{Validation of the Model}
In this section, numerical integration of the the two-fluid model
equations (\ref{Eqdeuxfluid1}-\ref{Eqdeuxfluid5}) are performed
using a pseudo-spectral code. Time marching is done
using second-order leapfrog finite difference scheme and even and odd
time-steps are periodically re-coupled by fourth-order
Runge-Kutta. The effective viscosity and diffusivity
are updated at each time step by resetting $E_{\rm th}=<Q>$.
The obtained data is compared with that directly produced from the truncated
Euler equation.
\begin{figure}[h!]
\begin{center}
  \includegraphics[width=0.48\textwidth,height=7cm]{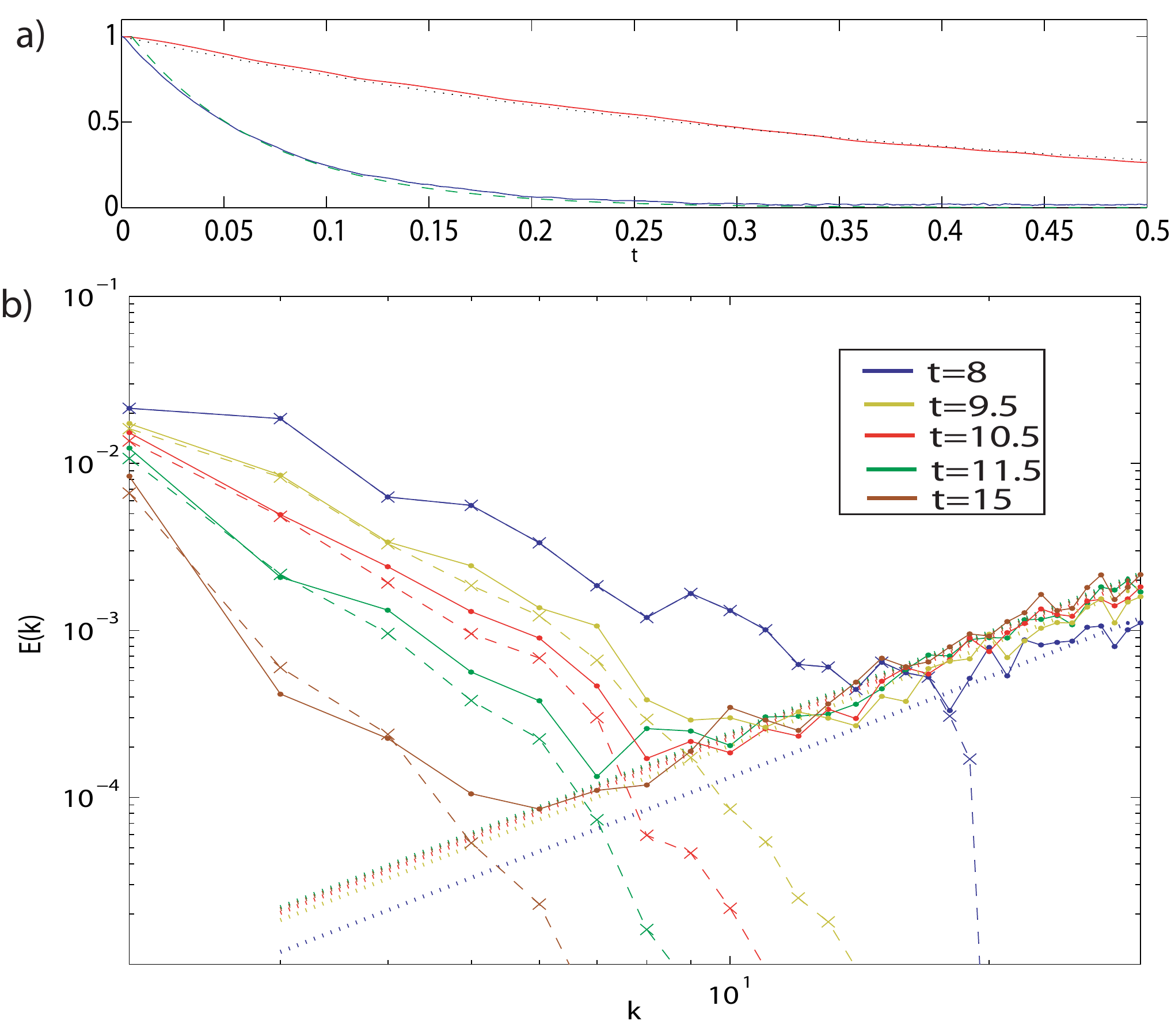}
  \caption{a)Time decay of rotation (\ref{rot+eq1}-\ref{rot+eq2}) (upper curve) and temperature modulation  (\ref{tempmod}) (bottom curve). Solid line: truncated Euler equations and dashed line: two-fluid model.
  b)Time-evolution of energy spectra, truncated Euler equation: solid lines and two-fluid model: dashed lines.}\label{FigDump}
\end{center}
\end{figure}

The time-evolutions resulting from initial data
(\ref{rot+eq1}-\ref{rot+eq2}) (in red) and (\ref{tempmod}) (in
blue), both normalized to one and with the same value of
$E_{\rm{th}}$ is displayed on figure \ref{FigDump}a . Good agreement
with the two-fluid model is obtained in both cases and the faster
relaxation of the temperature modulation is related to the smallness
of $P_{\rm{eff}}<1$.

We now compare, the evolution of non-trivial spectra of the truncated Euler
equation (\ref{eq_discrt}) and the two-fluid model. The
truncated Euler equation is integrated using the Taylor-Green
initial data. At $t\sim8$, when a clear scales separation is
present, the large-scale fields $\bf{v^<}$ (see eq.\ref{eq:Filt1})
and the heat $Q$ (eq.\ref{eq:Q}) are computed and used as initial
data for the two-fluid model
(\ref{Eqdeuxfluid1}-\ref{Eqdeuxfluid5}). The subsequent evolution of
the two-fluid model is then compared with that of the truncated
Euler equation.

Both spectra, plotted in figure \ref{FigDump}b, are in good
agreement. The straights lines represents the thermalized zone
$E(k,t)=c(t) k^2$ in the the spectrum of the truncated Euler
equation, where $c(t)$ is determined by the condition $\langle
Q(t)\rangle=\sum\limits_{k>k_{\rm{th}}}c(t) k^2$.

The value of $Q(t)$ and $\Delta Q/ Q$ are plotted in figure
\ref{FigTempVst} (solids lines); the evolution of the fluctuation of
the temperature are well reproduced too by the two-fluid model.
\section{Conclusion}
\label{sec:Conc}
The thermalized small scales were found
to follow a quasi-normal distribution.
The effective viscosity was determined,
using both EDQNM and Monte Carlo.
(Hypo)diffusion of heat was obtained and
the effective Prandtl number found to vanish
at small $k/k_{\rm max}$.
The two-fluid model was found to be in
good quantitative agreement with the original truncated Euler
equations.
%
We acknowledge useful scientific discussions with A. Pouquet and the
support of ECOS and CONICYT.
\bibliographystyle{unsrt}

\end{document}